\newcounter{myctr}
\def\myitem{\refstepcounter{myctr}\bibfont\noindent\ifnum\themyctr>9\else\phantom{0}\fi\hangindent17pt\themyctr.\enskip}

\documentclass{ws-ijqi}
\usepackage{hyperref}
\usepackage[super,sort,compress]{cite}

\usepackage{amsmath}
\usepackage{bbold}
\usepackage{braket}
\begin{document}

\newcommand*\per{\mathrm{perm}}
\catchline{}{}{}{}{}

\title{Which role does multiphoton interference play in small phase estimation in quantum Fourier transform interferometers?}

\author{OLAF ZIMMERMANN}

\address{Institut f{\"u}r Quantenoptik and Institut f{\"u}r Quantenphysik, Universit{\"a}t Ulm\\
D-89081 Ulm, Germany\\}

\author{VINCENZO TAMMA}

\address{Faculty of Science, SEES and Institute of Cosmology \& Gravitation, University of Portsmouth\\
Portsmouth, PO1 3QL, UK\footnote{Present address}\\
Institut f{\"u}r Quantenphysik and Center for Integrated Quantum Science and Technology (IQ\textsuperscript{ST}), Universit{\"a}t Ulm\\
D-89081 Ulm, Germany\\
vincenzo.tamma@port.ac.uk}
\maketitle


\begin{abstract}
Recently, quantum Fourier transform interferometers have been demonstrated to allow a quantum metrological enhancement in phase sensitivity for a small number $n$ of identical input single photons [\citen{QuFTInew,QuFTIold,OZBT}]. However, multiphoton distinguishability at the detectors can play an important role from an experimental point of view [\citen{Tamma}]. 
This raises a fundamental question: 
How is the phase sensitivity affected when the photons are completely distinguishable at the detectors and therefore do not interfere? 
In other words, which role does multiphoton interference play in these schemes? 
Here, we show that for small phase values the phase sensitivity achievable in the proposed schemes with indistinguishable photons is enhanced only by a constant factor with respect to the case of completely distinguishable photons at the detectors. Interestingly, this enhancement arises from the  interference of only a polynomial number (in $n$) of the total $n!$ multiphoton path amplitudes in the $n$-port interferometer. These results are independent of the number $n$ of single photons and of the phase weight factors employed at each interferometer channel.
\end{abstract}

\keywords{Multiphoton interference; Quantum metrology.}
\markboth{Olaf Zimmermann, Vincenzo Tamma}{Metrological advantage of multiphoton interference in quantum Fourier transform interferometers}

\section{Introduction}	

Multiphoton interference  [\citen{Tamma,RevModPhys.84.777}] has been demonstrated to be a powerful tool toward achieving quantum enhancement in parameter estimation   [\citen{You,Ge,YouRef3,YouRef1,YouRef4,YouRef6}].
Unfortunately, often the desired quantum metrological supremacy relies on the need of quantum states difficult to produce   [\citen{YouRef15,YouRef16,YouRef13,YouRef14}].   
An alternative approach relying only on single photon sources and a quantum Fourier transform interferometer (QuFTI) was recently proposed  in order to achieve a quantum metrological advantage in the case of small number n of single photons [\citen{QuFTInew,QuFTIold,You}]. Here,
we examine the 
role of multiphoton interference in the phase measurement accuracy of these techniques by considering the two cases of 
fully indistinguishable and fully distinguishable  input photons.
We show, that the accuracy is equal in both cases apart from a constant factor $\sqrt{2}$, that is independent of the number $n$ of input photons.
Therefore, multiphoton interference is able to provide only a constant advantage in phase measurement accuracy by using a QuFTI with $n$ input single photons, independently of the value of $n$.  
Our work generalizes the results obtained in Ref. [\citen{OZBT}], where the QuFTI  was first investigated only in the case of the fixed set of weight factors introduced in Ref. [\citen{QuFTIold}], to arbitrary phase weight factors (see Fig. 1).
The approach pursued here for single-parameter estimation can be also applied to estimate the role of multiphoton interference in multiparameter estimation schemes either with input single photon states or photon number states   [\citen{You,Ge,PhysRevLett.119.130504,YouRef1}].

\section{Phase accuracy for indistinguishable and distinguishable photons}

\begin{figure}[h]
\centerline{\includegraphics[width= \textwidth]{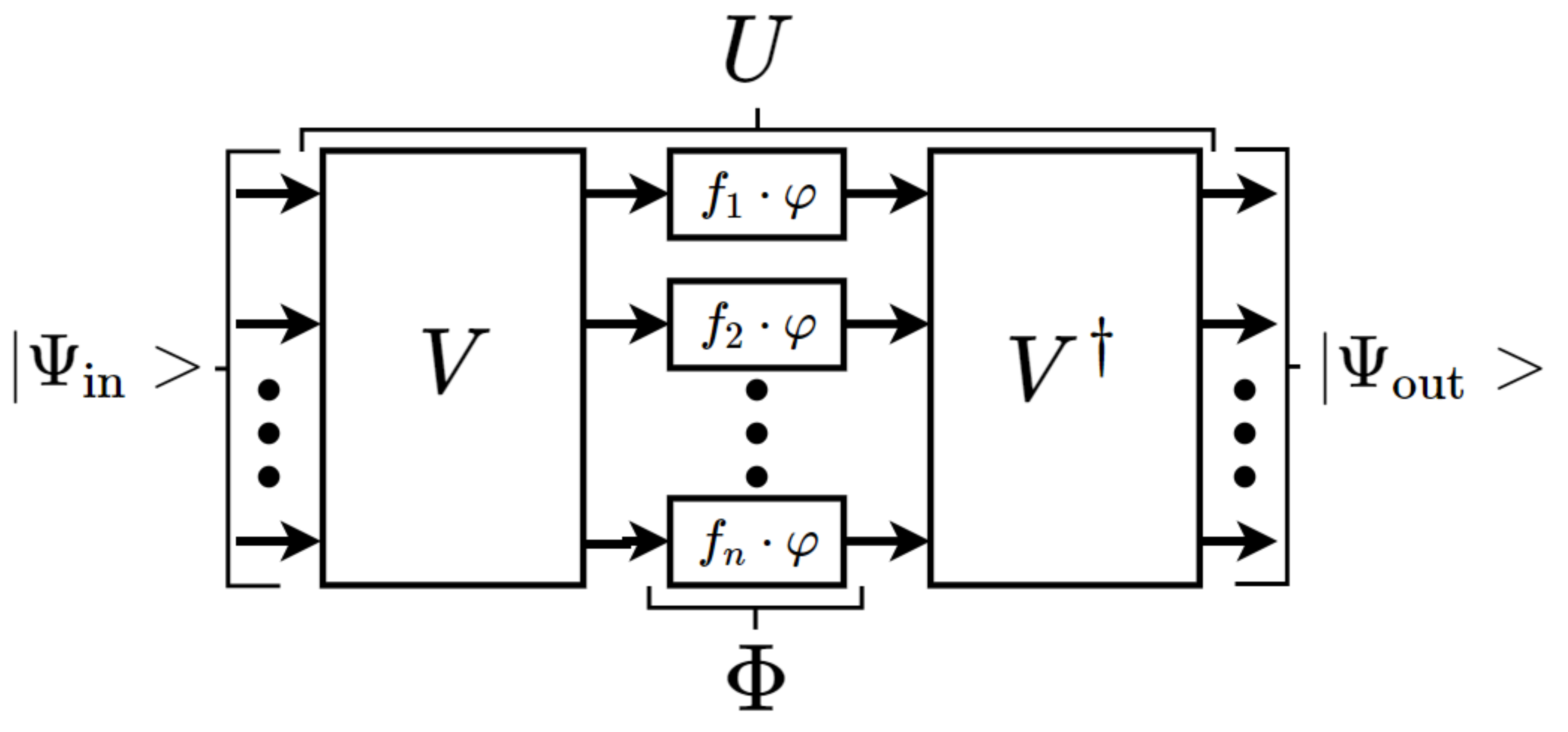}}
\vspace*{8pt}
\caption{The quantum Fourier transform interferometer  in Ref.   \citen{QuFTInew} is described by the unitary matrix ${U}={V}{\Phi}{V}^\dagger$. 
The interferometer is a succession of the $n$-mode quantum Fourier transform ${V}$, a phase shifter matrix ${\Phi}$, with an arbitrary set of phase weight factors $\{f_j\}$, $j=1,...,n,$ and the inverse $n$-mode quantum Fourier transform ${V}^\dagger$.\label{fig:QuFTI}}
\end{figure}

The QuFTI introduced in Ref.   \citen{QuFTInew} is an $n$-mode passive linear interferometer, determined by the succession of three components leading to the unitary matrix ${U}=V{\Phi}V^\dagger$ as illustrated in Fig. \ref{fig:QuFTI}.
The three components are the $n$-mode quantum Fourier transform, described by the matrix $V$ with elements
\begin{equation}
V_{j,k}=\frac{1}{\sqrt{n}} \mathrm{e}^{2\pi i \frac{(j-1)(k-1)}{n}},
\label{Vmatrix}
\end{equation}
with $j,k=1,...,n$, a general phase evolution, represented by the matrix $\Phi$ with elements
\begin{equation}
{\Phi}_{j,k}=\delta_{j,k}\mathrm{e}^{ i f_j \varphi}
\label{Phimatrix}
\end{equation} 
and the inverse quantum Fourier transform ${V}^\dagger$.
Here, $\varphi$ is the encoded phase to be measured, and $f_j$ is an arbitrary factor depending on the experimental scenario [\citen{QuFTInew}]. 
The interferometer matrix U is then defined by the elements 
\begin{equation}
U_{j,k}=\frac{1}{n}\sum\limits_{l=1}^n
	\omega_n^{(j-k)(l-1)}
	e^{i\varphi f_l},
\end{equation}
with $\omega_n=\exp(2\pi i/n)$, as derived in \ref{AppendixMatrix}.

Each interferometer port $j=1,...,n$, is fed with a single photon 
\begin{equation}
\ket{1[\xi_j]}_j:=\int \mathrm{d}\omega\, \xi_j(\omega)\,\hat{a}^\dagger(\omega)\ket{0}
\end{equation}
with spectrum $\xi_j$,
leading to the input state 
\begin{equation}
\ket{\Psi_{\mathrm{in}}}= \bigotimes_{j=1}^n \ket{1[\xi_j]}_j.
\end{equation}
The observable $\hat{O}$ that is measured to estimate the encoded phase $\varphi$ is the projection of the QuFTI output state $\ket{\Psi_{\mathrm{out}}}={U}\ket{\Psi_{\mathrm{in}}}$ onto the state
of one photon measured in each output mode $k\in\{1,...,n\}$ of the QuFTI, independently of the photonic inner modes (e.g. detection times and frequencies).
The phase accuracy is therefore determined by the probability $P= \braket{\hat{O}}=\braket{\hat{O}^2}$ of $n$-photon detections  as
\begin{equation}
\Delta\varphi
=\frac{\sqrt{\braket{\hat{O}^2}-\braket{\hat{O}}^2}}{\frac{\partial \braket{\hat{O}}}{\partial\varphi}}
=\frac{\sqrt{P-P^2}}{\frac{\mathrm{d}P}{\mathrm{d}\varphi}},
\label{PhaseSens}
\end{equation}
where the first equality is obtained by using standard error propagation.

The probability $P = P^{(I)}$ in the case of multiphoton interference arising from indistinguishable (I) photons (full overlap in the photonic spectra, $\int \xi_j\xi_k\mathrm{d}\omega=1$ $\forall j,k$) is given by [\citen{Tamma,Aaronson,Scheel}]
\begin{equation}
P^{(I)}=|\per( U)|^2
\label{perU}
\end{equation}
where the permanent of a matrix $\Gamma$ is defined as 
\begin{align}
\per(\Gamma):=\sum\limits_{\sigma\in S_n}\prod_{j=1}^n \Gamma_{j,\sigma(j)}.
\label{perdef}
\end{align}
The permanent in Eq. \eqref{perU} is given by the superposition of $n!$ $n$-photon amplitudes corresponding to the $n!$ possible ways for the $n$ photons to trigger the $n$ detectors. 
Each multiphoton amplitude is identified by a permutation $\sigma(j)$ of the output ports where the corresponding photons at each input port  $j=1,...,n$ can exit. 

In the case of completely distinguishable (D) input photons (no overlap in the photonic spectra, $\int\xi_j\xi_k\mathrm{d}\omega=0 $ $\forall j\neq k$) the probability $P= P^{(D)}$ is given by [\citen{Tamma}]
\begin{equation}
P^{(D)}=\per( T),
\label{perT}
\end{equation}
with the matrix $T$ defined by the elements 
$T_{j,k}=|U_{j,k}|^2$. Here, all multiphoton interference terms in Eq. (7) vanished due to the distinguishability of the n photons at the detectors.

\subsection{Phase accuracy for small phase values}

In this section we calculate the phase accuracy $\Delta\varphi$ in the case of fully indistinguishable and fully distinguishable photons.
We restrict our discussion to small values of the phase $\varphi$, where higher values of accuracy can be achieved [\citen{OZBT,QuFTInew}].\\

\begin{figure}[h]
\centerline{\includegraphics[width= \textwidth]{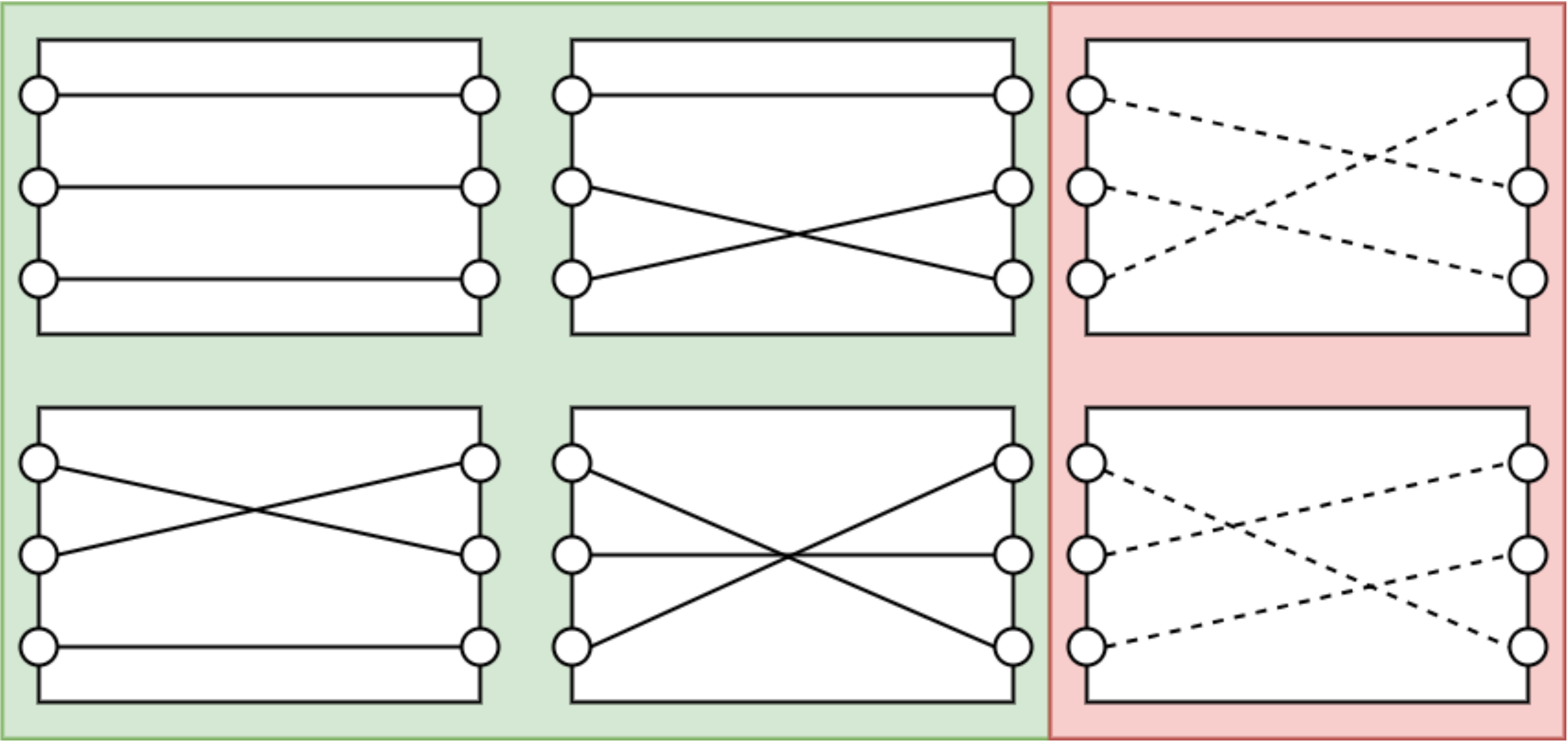}}
\vspace*{8pt}
\caption{
Interference of n-photon amplitudes in a QuFTI depicted, as an example, for n=3.
Only $n$-photon amplitudes associated with an exchange of two photons at the output ports of the QuFTI with respect to the identity permutation at the top left contribute to the multiphoton interference with terms of the order $\varphi^2$. 
These amplitudes, in addition to the identity amplitude, are illustrated here with green background and continuous-line paths.
The remaining n-photon amplitudes, indicated with a red background and dashed-line paths, are suppressed since they contribute to higher order terms in $\varphi$. \label{fig:PhotonPermutations}}
\end{figure}

We first address the case of indistinguishable photons.
By considering only terms of the order of $\varphi^2$, Eq. \ref{perU} can be rewritten as
\begin{align}
P^{(I)}=&\left|\left(1+\frac{i}{n}\varphi A	
		-\frac{1}{2n}\varphi^2 B
		\right)^{n}
		+
		\sum\limits_{\sigma_2}\prod_{j\neq\sigma_2(j)}
		\left(\frac{i\varphi}{n}C_{j,\sigma_2(j)}
		-\frac{\varphi^2}{2n}D_{j,\sigma_2(j)}\right)\right|^2,
		\label{PermanentUT}
\end{align}
as derived in \ref{AppendixPFO},
where $A:=\sum_j f_j$, $B:=\sum_j f_j^2$,
$C_{j,k}:=\sum\limits_{l}\omega_n^{(j-k)(l-1)}f_l$, $D_{j,k}:=\sum\limits_{l}\omega_n^{(j-k)(l-1)}f_l^2$ and ${\sigma_m}$ are the permutations interchanging $m$ elements.
Interestingly, only the identity permutation $\sigma_0$ (first term in Eq. \eqref{PermanentUT}) and the $n(n-1)/2$ permutations 
$\{\sigma_2\}$ interchanging two output ports contribute to the multiphoton interference. 
As an example, these permutations are illustrated in Fig. \ref{fig:PhotonPermutations} in the case of $n=3$.\\

On the other hand, in the case of fully distinguishable photons and neglecting terms of order $\varphi^2$ and higher, Eq. \eqref{perT} reduces to
\begin{align}
P^{(D)}=&\left|1+\frac{i}{n}\varphi A	
		-\frac{1}{2n}\varphi^2 B
		\right|^{2n},
\label{ProbT}
\end{align}
 corresponding to the modulus square of the first multiphoton amplitude in Eq. \eqref{PermanentUT} associated with the identity permutation.\\
 
We conclude that the difference in the probability for indistinguishable photons with respect to distinguishable photons only arises from the contribution of the $n$-photon amplitudes associated with an exchange of two photons at the output ports of the QuFTI.\\

In \ref{AppendixPFO} and \ref{AppendixPNO} we find that the probabilities in Eq. \eqref{PermanentUT} and Eq. \eqref{ProbT} reduce, respectively, to 
\begin{equation}
P^{(I)}=1-2\varphi^2(B-\frac{A^2}{n})
\label{P1}
\end{equation}
and 
\begin{equation}
P^{(D)}=1-\varphi^2(B-\frac{A^2}{n}).
\label{P2}
\end{equation}
By using the results in Eqs. \eqref{P1} and \eqref{P2} the phase accuracy in Eq. \eqref{PhaseSens} becomes
\begin{equation}
\Delta\varphi^{(I)}=\sqrt{\frac{n}{8(Bn-A)}}
\end{equation}
for completely indistinguishable photons
 and
 \begin{equation}
 \Delta\varphi^{(D)}=\sqrt{\frac{n}{4(Bn-A)}}
\end{equation}
for completely distinguishable photons.
Thus, the interferometric phase sensitivities given fully indistinguishable and fully distinguishable input photons  only differ by a constant factor of $\sqrt{2}$. 
This factor is independent of the number $n$ of input photons and of the phase weight factors $f_j$.

\section{Conclusions}

We demonstrated that a constant factor $\sqrt{2}$ represents the only enhancement in phase sensitivity  achievable by the QuFTI for fully indistinguishable input single photons in comparison to the case of fully distinguishable input single photons.
These results hold for arbitrary numbers $n$ of single photons injected in the interferometer, and for arbitrary phase factors $f_j$ introduced in each interferometer channel.
This constant quantum enhancement arises from the contribution of interfering multiphoton amplitudes associated with all the possible ways to exchange only two photons at the output ports (see Fig. \ref{fig:PhotonPermutations}).  Interestingly, these amplitudes, together with the contribution where no exchange occurs, represent only $1+n(n-1)/2$ of $n!$ interfering amplitudes which would contribute, in general, for larger values of the phase $\varphi$ in an n-port interferometer. 
This motivates further investigations of how to maximize the quantum metrological enhancement achievable by $n$-photon interference in linear optics quantum metrology depending on the values of the phases to be measured and on the interferometer configuration [\citen{OZBT,Ge,You}].
Furthermore, this work can stimulate further studies about the role of multiphoton interference in quantum information processing [\citen{Aaronson,Tamma2,Tamma3,Tamma5,Tamma4}].

\section*{Acknowledgements}

V.T. is thankful for discussions with M. Foss-Feig, W. Ge and K. Jacobs.
Research was partially sponsored by the Army Research Laboratory and was accomplished under Cooperative Agreement Number W911NF-17-2-0179. 
The views and conclusions contained in this document are those of the authors and should not be interpreted as representing the official policies, either expressed or implied, of the Army Research Laboratory or the U.S. Government. 
The U.S. Government is authorized to reproduce and distribute reprints for Government purposes notwithstanding any copyright notation herein.
V. T. and O. Z. are also very thankful to W. P. Schleich for the time passed at the Institute of Quantum Physics in Ulm, where some of the ideas behind this work were first developed [\citen{OZBT,Tamma}].

\renewcommand\bibname{References}

\appendix
\section{QuFTI Matrix elements for small phases $\varphi$ \label{AppendixMatrix}}

We derive the matrix elements $U_{j,k}$, with $j,k=1,...,n$ for the matrix $U= V\Phi V^\dagger$ associated with an $n$-port QuFTI by using Eqs. \eqref{Vmatrix} and \eqref{Phimatrix}:
\begin{align}
U_{j,k}:=&\sum\limits_{l,m=1}^n
	V_{j,l}\Phi_{l,m}V_{m,k}^\dagger \nonumber
=\frac{1}{n}\sum\limits_{l,m=1}^n
	\omega_n^{(j-1)(l-1)}
	\underbrace{\delta_{l,m}e^{i\varphi f_l}}_{\Phi_{l,m}}
	\omega_n^{(m-1)(1-k)}\\
=&\frac{1}{n}\sum\limits_{l=1}^n
	\omega_n^{(j-1)(l-1)}
	e^{i\varphi f_l}
	\omega_n^{(l-1)(1-k)}
=\frac{1}{n}\sum\limits_{l=1}^n
	\omega_n^{(j-k)(l-1)}
	e^{i\varphi f_l}.
\end{align}
Here, $\omega_n=\exp(2\pi i/n)$ is the $n$\textsuperscript{th} root of unity and $f_j$ is an arbitrary factor, dependent on the channel $j$.
We expand the exponential for small phases $\varphi$ and obtain
\begin{align}
U_{j,k}=&\frac{1}{n}\sum\limits_{l=1}^n
	\omega_n^{(j-k)(l-1)}
	\left( 1+i\varphi f_l-\frac{1}{2}\varphi^2f_l^2 \right) \nonumber\\
=&\begin{cases}
		1+\frac{i}{n}\varphi \underbrace{\sum\limits_{l=1}^n f_l}_{=:A}	
		-\frac{1}{2n}\varphi^2 \underbrace{\sum\limits_{l=1}^n f_l^2}_{=:B}	 &\mbox{, } j= k\\
		\underbrace{ \frac{1}{n}\sum\limits_{l=1}^n\omega_n^{(j-k)(l-1)}}_{=0}
		+\frac{i\varphi}{n}\underbrace{\sum\limits_{l=1}^n\omega_n^{(j-k)(l-1)}f_l}_{=:C_{j,k}}
		-\frac{\varphi^2}{2n}\underbrace{\sum\limits_{l=1}^n\omega_n^{(j-k)(l-1)}f_l^2}_{=:D_{j,k}}		&\mbox{, } j\neq k\\
	\end{cases}\nonumber\\
	=&\begin{cases}
		1+\frac{i}{n}\varphi A	
		-\frac{1}{2n}\varphi^2 B	 &\mbox{, } j= k\\
		\frac{i\varphi}{n}C_{j,k}
		-\frac{\varphi^2}{2n}D_{j,k}		&\mbox{, } j\neq k.\\
	\end{cases}
	\label{matrixElementsU}
\end{align}

\section{$P^{(I)}$ for small phase values $\varphi$ \label{AppendixPFO}}
We calculate the probability $P^{(I)}=|\per(U)|^2$ for an $n$-fold single photon detection given an input of indistinguishable single photons from the permanent of the QuFTI matrix $U$ with elements defined by Eq. \eqref{matrixElementsU}. 
By employing the general definition of the permanent in Eq. \eqref{perdef}
and defining the permutations $\sigma_m$ interchanging $m$ elements we obtain
\begin{align}
\per(U)=&\left(1+\frac{i}{n}\varphi A	
		-\frac{1}{2n}\varphi^2 B\right)^{n}
		\nonumber\\&+
		\sum\limits_{m=2}^n\left(1+\frac{i}{n}\varphi A	
		-\frac{1}{2n}\varphi^2 B\right)^{n-m} 
		\cdot\underbrace{\sum\limits_{\sigma_m}\prod_{j\neq\sigma_m(j)}
		\left(\frac{i\varphi}{n}C_{j,\sigma_m(j)}
		-\frac{\varphi^2}{2n}D_{j,\sigma_m(j)}\right)}_{O\left(\varphi^m\right)}.
		\label{PermanentU}
\end{align}
Given that the terms in the inner sum are of the order $\varphi^m$ we can assume that it is sufficient to consider only permutations $\sigma_m$ with $m\leq 2$ for small phase values $\varphi$. 
Since no permutation can interchange only 1 element, the permanent in Eq. \eqref{PermanentU} is given by the contribution of the identity $\sigma_0$
\begin{align}
T_{\sigma_0}:=&\left(1+\frac{i}{n}\varphi A	
		-\frac{1}{2n}\varphi^2 B\right)^{n}=1+i\varphi A	
		-\frac{1}{2}\varphi^2 B-\frac{n-1}{2n }\varphi^2 A^2
		\label{IDcontribution}
\end{align}
and the sum 
\begin{align}
T_{\{\sigma_2\}} :=&\left(1+\frac{i}{n}\varphi A	
		-\frac{1}{2n}\varphi^2 B\right)^{n-2} 
		\frac{1}{2}\sum\limits_{j_1=1}^n\sum\limits_{\stackrel{j_2=1}{j_2\neq j_1}}^n
		\left(\frac{i\varphi}{n}C_{j_1,{j_2}}
		-\frac{\varphi^2}{2n}D_{j_1,{j_2}}\right)
		\nonumber\\&\cdot\left(\frac{i\varphi}{n}C_{{j_2},j_1}
		-\frac{\varphi^2}{2n}D_{{j_2},j_1}\right)\nonumber\\
=&-\frac{\varphi^2}{n^2} 
\frac{1}{2}\sum\limits_{j_1=1}^n\sum\limits_{\stackrel{j_2=1}{j_2\neq j_1}}^n
C_{j_1,{j_2}}C_{{j_2},j_1}+O(\varphi^3)
		\label{S2contribution}
\end{align}
of the contributions from all $n(n-1)/2$ permutations $\sigma_2$ that interchange only $2$ elements, labelled as $j_1,j_2$. Here, the factor $1/2$ arises from the fact that each permutation is considered twice in the sum.
By considering only contributions up to the order $\varphi^2$,  Eq. \eqref{S2contribution} becomes
\begin{align}
T_{\{\sigma_2\}}=&-\frac{\varphi^2}{2n^2}\sum\limits_{j_1=1}^n\sum\limits_{\stackrel{j_2=1}{j_2\neq j_1}}^n 
	C_{j_1,j_2}C_{j_2,j_1}
=-\frac{\varphi^2}{2n^2}\sum\limits_{l_1=1}^n\sum\limits_{l_2=1}^n	
	f_{l_1}f_{l_2}
	\sum\limits_{j_1=1}^n\sum\limits_{\stackrel{j_2=1}{j_2\neq j_1}}^n
	\omega_n^{(j_1-j_2)(l_1-l_2)}	\nonumber\\
=&-\frac{\varphi^2}{2n^2}\sum\limits_{l_1=1}^n\sum\limits_{l_2=1}^n	
	f_{l_1}f_{l_2}	
	\sum\limits_{j_1=1}^n\sum\limits_{\stackrel{j=1-j_1}{j\neq 0}}^{n-j_1}
	\omega_n^{-j(l_1-l_2)} \nonumber\\
=&-\frac{\varphi^2}{2n^2}\sum\limits_{l_1=1}^n\sum\limits_{l_2=1}^n	
	f_{l_1}f_{l_2}	
	\sum\limits_{j_1=1}^n \left( \sum\limits_{j=1-j_1}^{n-j_1}
	\omega_n^{-j(l_1-l_2)}-1\right)\nonumber\\
=&-\frac{\varphi^2}{2n^2}\sum\limits_{l_1=1}^n\sum\limits_{l_2=1}^n	
	f_{l_1}f_{l_2}	
	\sum\limits_{j_1=1}^n \left( \delta_{l_1,l_2}n-1\right)
=-\frac{\varphi^2}{2n^2}\sum\limits_{l_1=1}^n\sum\limits_{l_2=1}^n	
	f_{l_1}f_{l_2}n \left( \delta_{l_1,l_2}n-1\right)\nonumber\\
=&-\frac{\varphi^2}{2}B+\frac{\varphi^2}{2n}A^2
	.
		\label{S2sum}
\end{align}
By summing the contributions in Eq. \eqref{IDcontribution} and Eq. \eqref{S2sum}, Eq. \eqref{PermanentU} becomes
\begin{align}
\per(U)=T_{\sigma_0}+T_{\{\sigma_2\}}=1+i\varphi A-\varphi^2 B-\frac{n-2}{2n}\varphi^2 A^2.
\end{align}
From Eq. \eqref{perU}, this leads to the expression
\begin{align}
P^{(I)}=1+\varphi^2 A^2-2\varphi^2B-\frac{n-2}{n}\varphi^2A^2
=1-2\varphi^2(B-\frac{A^2}{n}).
\label{APPENDIXProbU}
\end{align}

\section{$P^{(D)}$ for small phase values $\varphi$ \label{AppendixPNO}}

We calculate the probability $P^{(D)}=\per(T)$ 
in Eq. \eqref{perT}, where the elements $T_{j,k}=|U_{j,k}|^2$ of the matrix $T$ follow directly from Eq.\eqref{matrixElementsU} as 
\begin{align}
T_{j,k}=&\left|U_{j,k}\right|^2
=\begin{cases}
		1+\frac{1}{n^2}\varphi^2 A^2	
		-\frac{1}{n}\varphi^2 B &\mbox{, } j= k\\
		-\frac{\varphi^2}{n^2}\left| C_{j,k} \right|^2
		&\mbox{, } j\neq k.\\
	\end{cases}
	\label{matrixElementsT}
\end{align}
By using the definition of permanent in Eq. \eqref{perdef}, Eq. \eqref{perT} becomes
\begin{align}
P^{(D)}=&\left(1+\frac{1}{n^2}\varphi^2 A^2	
		-\frac{1}{n}\varphi^2 B \right)^n
		\nonumber\\&
	+	
	\sum\limits_{m=2}^n\left(1+\frac{1}{n^2}\varphi^2 A^2	
		-\frac{1}{n}\varphi^2 B \right)^{(n-m)}
				\underbrace{\sum\limits_{\sigma_m}\prod_{j\neq\sigma_m(j)}
		\left(-\frac{\varphi^2}{n^2}\left| C_{j,k} \right|^2
		\right)}_{O\left(\varphi^{2m}\right)}.
\label{APPENDIXProbT}
\end{align}
We point out that, given the dependence on $\varphi^2$ of the non-diagonal elements in Eq. \eqref{matrixElementsT}, the contributions from the permutations $\sigma_m$ in Eq. \eqref{APPENDIXProbT} are of order $\varphi^{2m}$. Therefore, by neglecting terms of higher order than $\varphi^2$,
Eq. \eqref{APPENDIXProbT} reduces to its first term associated with the identity permutation as
\begin{equation}
P^{(D)}=1-\varphi^2(B-\frac{A^2}{n}),
\end{equation}
which corresponds to 
\begin{equation}
P^{(D)} = \left|T_{\sigma_0}\right|^2,
\end{equation}
with $T_{\sigma_0}$ defined in Eq. (B2).


\end{document}